# Thermally driven continuous-wave and pulsed optical vortex

Yitian Ding,[1,2] Miaomiao Xu,[1,3] Yongguang Zhao,[1] Haohai Yu,[1*] Huaijin Zhang,[1] Zhengping Wang,[1] and Jiyang Wang[1]

[1]*State Key Laboratory of Crystal Materials and Institute of Crystal Materials, Shandong University, Jinan 250100, China*
[2]*Department of physics, Shandong University, Jinan 250100, China*
[3]*Taishan College, Shandong University, Jinan 250100, China*
**Corresponding author: haohaiyu@sdu.edu.cn*

We demonstrated the continuous-wave (cw) and pulsed optical vortex with topological charges driven by heat generated during the lasing process without introducing the astigmatism effect and reducing lasing efficiency. During the lasing process, the topological charges were changeable by the thermal-induced lens and selected by the mode-matching between the pump and oscillating beams. With a graphene sample as the saturable absorber, the pulsed optical vortex was achieved at the wavelength of 1.36 μm, which identified that graphene could be used as a pulse modulator for the generation of pulsed optical vortex. It could be believed that the thermally driven cw and pulsed optical vortex should have various promising applications based on the compact structure, changeable topological charges and specific wavelength.© 2013 Optical Society of America

Angular momentum (AM) is a universal physical concept and can be carried by a rotated particle. The orbital and spin are two types of AM. Similar with rotated particles such as electrons, light can also have AM when it has a helical phase or rotated polarization direction [1, 2]. Both of them can transfer torque [3, 4]. However, the spin AM per photon has only three values 0 and $\pm\hbar$ corresponding to the linear and circular polarization with different rotated directions, and the orbital AM per photon $l\hbar$ is generated by the helical phase and determined by the topological charge or azimuthal mode index $l$ [2]. Up to now, their generation [4-6], propagation dynamics [7, 8], entanglement [9], interaction with matter [10], etc., have been drawing much attention over the past few decades, since they belong to the wave-particle duality [2] and reveal the nature of the optical angular momentum [11]. Laguerre-Gaussian ($LG_{p,l}$) laser modes are a type of eigen modes of the laser cavity and carry orbital angular momentum equal to $l\hbar$ per photon. The previous investigation has identified that the astigmatism effect is essential for the generation of high-order eigen laser modes, no matter for the Hermite–Gaussian ($HG_{p,l}$) or $LG_{p,l}$ laser modes [12]. However, the introduction of astigmatism effects into the laser cavity would reduce the lasing efficiency since the astigmatism would reduce the mode-matching degree between pump and oscillating beams [13].

In the lasing process, the heat is unavoidable and mainly caused by the quantum defects [14]. The thermal effects generated are considered to be deleterious on the laser output performance, especially in the high-power lasers. Thermal induced lens is the direct result of thermal effects whose focal length is tunable by the absorbed pump power [15] and change the size of oscillating modes. The size of oscillating high-order eigen modes of the laser cavity is proportional to their order and sizes of the fundamental modes determined by the laser cavity. By mode-matching between the pump and oscillating beams, the order of achieved laser modes can be driven thermally and no obvious reduction of the optical efficiency is generated. In the universal neodymium doped crystal lasers, the most serious thermal effects are generated during the lasing process at the wavelength of 1.3 μm, since the quantum defect are the largest. The lasers at the wavelength about 1.3 μm have great applications in the specific fields of medical treatment, optical fiber communication, efficient production of red radiation by frequency doubling, etc. [16], aside from the absorption of some specific matter at this wavelength. In this letter, we demonstrate the thermally driven continuous-wave (cw) and pulsed $LG_{p,l}$ laser modes at the wavelength of 1.3 μm with tunable topological charges. The results also identified that graphene can be used as a saturable absorber for the generation of pulsed optical vortex.

The pump source was a fiber-coupled laser diode (LD) with a central wavelength around 808 nm. The core size of the fiber was 100 μm in radius, and the numerical aperture was 0.22. The output intensity from the fiber distributes as a doughnut shape. The symmetry of the doughnut shape distribution determined that the laser system could be described with the cylindrical coordinate which meet the requirement of the generation of LG eigen modes of a cavity. Using an imaging unit with a beam compression ratio of 1:1, the pump light was focused into the crystal with the core size of 100μm in radius. The experimental configuration used for the generation of cw and pulsed optical vortex was shown in Fig.1. The gain medium was Nd doped $(Lu_{0.5}Y_{0.5})_2SiO_5$ (Nd:LYSO) crystal, cut a long its c axis and with a size of 3 mm×3mm×10 mm, whose end faces were polished and uncoated. The crystal was wrapped with indium foil and mounted in a water-cooled copper block with cooling temperature of 20 °C. The emission peak with the longest wavelength was located at about 1.36 μm [17], which determined the quantum defect was 40.5% during lasing process in this wavelength band. A plane-concave resonant cavity was used to run the laser. A concave mirror (M1) with a curvature radius of 200 mm, which was high-transmission (HT) coated at 808 nm and 1.06 μm to 1.08 μm, and highly refractive at 1.3 μm to 1.4 μm, was used as the input mirror. A flat mirror (M2) was the output coupler, which was also HT coated at 1.06μm to 1.08 μm with an optimized transmission of 5% at 1.3μm to 1.4 μm. The oscillation at 1.06 μm to 1.08 μm in the cavity was suppressed by the HT coating at this wavelength band. The length of the cavity was about 20 mm. For the pulsed lasers, a multi-layered graphene with the initial transmission of 87% was used as the saturable absorber which was inset between the gain material and output coupler.

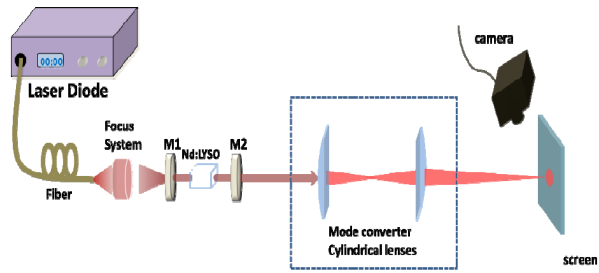

Fig. 1. Experimental configuration of cw and pulsed optical vortex

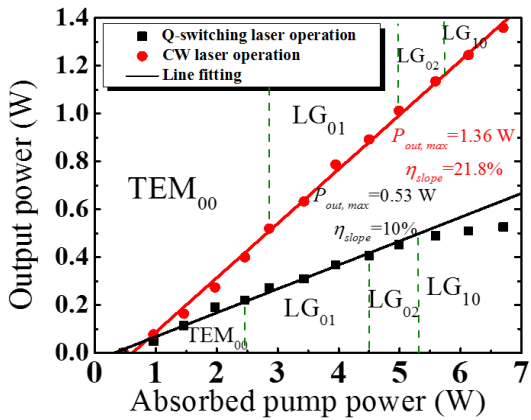

Fig. 2. Dependence of the cw and pulsed output power of $LG_{p,l}$ modes on the absorbed pump power.

The output power performance was shown in Fig. 2. The threshold was measured to be 0.45 W and the maximum output power was 1.36 W under the absorbed pump power of 6.7 W with the optical conversion and slope efficiency of 20.3% and 21.8%, respectively. The achieved typical transverse patterns were also shown in this figure, corresponding to the output power and absorbed pump power. From this figure, it could be found that the output transverse modes belong to $LG_{p,l}$ with the topological charge of $l\hbar$. Based on the ABCD matrix, the oscillating modes was infinite if there was no thermal lens generated in the crystal, since the cavity used was a concave-plano one. The $LG_{0,l}$ mode size could be approximately shown as $\omega_0(l+1)^{1/2}$ [18, 19], where $\omega_0$ was the fundamental mode size which can be calculated by the ABCD matrix. With the method presented by Song et. al [20], the thermal focal length could be measured with this cavity. Based on the calculation and measurement of the thermal focal length, it could be found that, in the used laser cavity, the $LG_{0,l}$ mode size was reduced as the decrease of the thermal induced focal length if the focal length is larger than about 3 mm. The modes with $l$=1 well matched with the pump beam in the pump range from 2.86 to 4.99 W, however, the modes with $l$=2 was well matched from 4.99 to 5.7 W. Increase the pump power over than 5.7 W, the $p$=1 mode appeared which was shown in the inset of Fig. 5. Considering the large quantum defects (40.5%), relatively poor thermal properties of the used crystal [21] and its low absorption efficiency (30.8%) at the pump beam, we did not increase the pump power further, so as to avoid the cracking. Therefore, it could be concluded that the generated optical vortex was thermally driven and the mode-matching was the selection rule for the order and topological charge of $LG_{0,l}$ modes. To identifying the $LG_{p,l}$ modes driven thermally, the cooling temperature was promoted to 25 °C which generated the higher $LG_{0,l}$ mode threshold (over than 3 W), since the thermal focal length is determined by the difference in temperature between the cooling sides on and the pump core in the crystal [15].

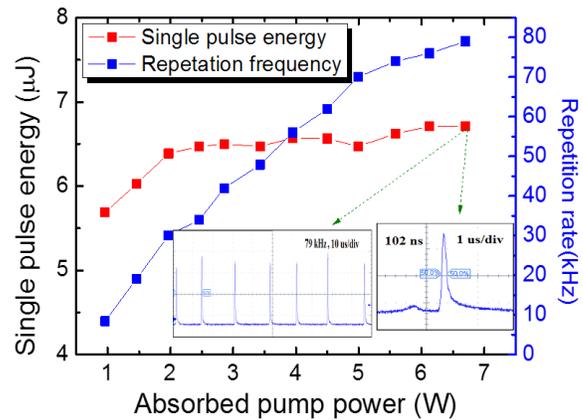

Fig. 3. Dependence of the repetition rate and pulse energy on the absorbed pump power. Inset: Pulse profile with width of 102 ns and repetition rate of 79 kHz.

To confirm the topological charge of achieved $LG_{0,l}$ modes, a mode converter made up of two identical cylindrical lenses was used to transform the $LG_{0,l}$ to

$HG_{0,n}$ modes, which could introduce a $i^k$ to the decomposition of $LG_{0,l}$ [2]. The plain surfaces of the two lenses were set to be paralleled to each other with the axis of the cylinder pointing vertically. In the experiment, we prepared the distance between the two cylindrical lenses to be precisely $\sqrt{2}\,f$, where $f$ was the focal length of the two lenses. Figure.4 showed the transverse patterns of the $LG_{0,l}$ modes and converted $HG_{0,l}$ modes. From this figure, we could identify the achieved $LG_{0,l}$ modes possessing the topological charges of $l=1$ and 2. In all the lasing process, there was no astigmatism effect in the lasing cavity and the slope efficiency of 21.8% was normal for the quantum defects of 40.5%. We believed that the efficiency could be improved if the laser crystal was suitable coated at the lasing wavelength.

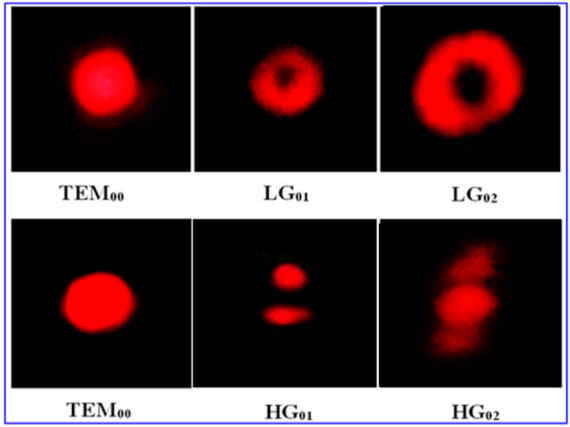

Fig. 4. The transverse pattern of the laser beam. Upper line: the achieved $LG_{0,l}$ modes. Down line: the converted $HG_{0,l}$ modes corresponding the $LG_{0,l}$ modes.

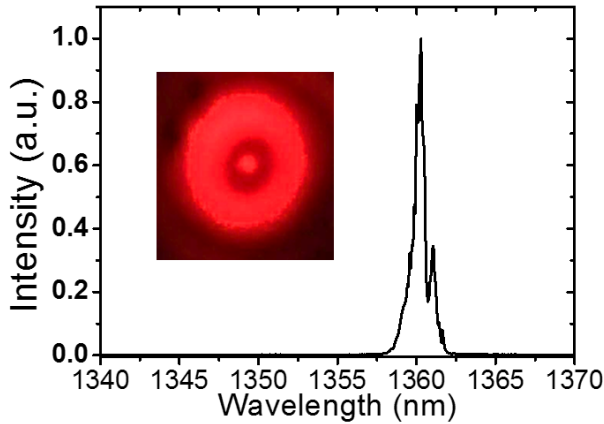

Fig.5. Laser spectrum of cw and pulsed vortex. Inset: The transverse pattern of the laser beam with $LG_{1,0}$ mode

Graphene has been investigated as a universal saturable absorber for the generation of pulsed lasers. By inserting the graphene into the cavity, the optical pulses were generated. The passively Q-switched laser performance was also presented in Fig. 2. The threshold was 0.47 W, a bit larger than the cw one. The highest output power was 0.53 W under the absorbed pump power of 6.7 W, with the slope efficiency of 8.5%. The repetition rate and pulse width were recorded with a DPO7104 digital oscilloscope (1 GHz bandwidth and 10 Gs/s sampling rate, Tektronix Inc.). As showed in Fig. 3, the repetition rate increased from 8.4 kHz to 79 kHz and the pulse energy increased from 5.69 µJ to 6.71 µJ in all pump power range. The shortest pulse was 102 ns under the pump power of 6.7 W. The pulse profile with the width of 102 ns and typical repetition rate of 79 kHz were also presented in the inset of Fig. 3. It should be noted that the thresholds of $LG_{0,1}$, $LG_{0,2}$ and $LG_{1,0}$ modes were respective 2.46 W 4.5 W and 5.3 W, obviously smaller than the corresponding cw thresholds. The observed $LG_{0,1}$, $LG_{0,2}$ and $LG_{1,0}$ modes are almost similar with those shown in 4. Considering the peak power of the output pulses, the intracavity peak intensity of the pulsed lasers was about 48 times larger than the cw one, which indicated that the nonlinear refractive index effects in the pulsed regime should be responsible for the lower threshold of high-order modes. From the pulsed results, no topological charge was found to be lost, which identified that there was no angular momentum transfer between the graphene and vortex pulses and the graphene could be used as a saturable absorber for the vortex pulses. Using an optical spectrum analyzer, the laser spectrum was measured to be located at about 1.36 µm, which is shown in Fig. 5.

In conclusion, thermally driven cw and pulsed optical vortex are directly generated by mode-matching selection. Considering the large quantum defects and specific application, the optical vortex at the wavelength of 1.36 µm are demonstrated. The results also identified that graphene can be used as a pulse modulator. We believe that the generated cw and pulsed optical vortex should find some promising applications in quantum optics, investigation of the interaction of orbital angular momentum of photons and matter, nonlinear optics, optical communications, etc.

This work is supported by the National Natural Science Foundation of China (Nos. 51025210, 51102156 and 51272131), and Shangdong Province Young and Middle-Aged Scientists Research Awards Fund (BS2011CL024).
Y. Ding and M. Xu contributed equally to this work